\documentclass[12pt,letterpaper]{article}
\usepackage{osajnl2}
\usepackage{amsmath}
\usepackage{pb-diagram}
\usepackage{hyperref}
\usepackage{multirow}
\usepackage{graphicx}
\usepackage{txfonts}
\usepackage{verbatim}
\newcommand{\aaps}{Astronomy and Astrophysics Supplement}
\newcommand{\apss}{Astrophysics and Space Science}

\begin{document}

\title{Interval estimate with probabilistic background constraints in deconvolution}

\author{Zhuo-xi Huo$^{1,2,3}$ and Jian-feng Zhou$^{1,2,3,*}$}
\address{$^1$Department of Engineering Physics and Center for Astrophysics, Tsinghua University, Beijing 100084, China}
\address{$^2$Key Laboratory of Particle \& Radiation Imaging (Tsinghua University), Ministry of Education, Beijing 100084, China}
\address{$^3$Key Laboratory of High Energy Radiation Imaging Fundamental Science for National Defense, Beijing 100084, China}
\address{$^*$Corresponding author: zhoujf@mail.tsinghua.edu.cn}

\begin{abstract}
We present in this article the use of probabilistic background constraints in astronomical image deconvolution to 
approach to a solution as an interval estimate. 
We elaborate our objective -- the interval estimate of the unknown object from observed data and our 
approach -- monte-carlo experiment and analysis of marginal distributions of image values. 
One-dimensional observation and deconvolution using proposed approach are simulated. 
Confidence intervals reveal the uncertainties due to the background constraint are calculated and significance levels for sources retrieved from restored images are provided.
\end{abstract}

\ocis{100.0100, 100.1830.}

\maketitle


\section{Introduction}
\subsection{Image restoration and deconvolution}
An imaging system receives the object intensity distribution $O$ as input while yields observed intensity distribution $I$ as its output. In most cases (more specifically, if the imaging system is approximately linear and shift-invariant, and the output data is contaminated by independent additive noise) such a system can be described exactly by the following equation:
\begin{equation}
\begin{split}
I(x) & = \int_{-\infty}^\infty O(x_1) P(x - x_1) \mathrm{d} x_1 + N(x) \\
        & = \left(O \ast P\right)(x) + N(x)
\end{split}
\label{eq-conv}
\end{equation}
where $x$ is a spatial coordinate variable, $P$ is the point spread function (PSF), $N$ is the additive noise and the asterisk $\ast$ denotes the convolution operation\cite{meister2009}.

The problem of reversing the effects of convolution on the observed data $I$, namely, restore the object intensity $O$, with an observation $I$, as well as a determinate PSF $P$ is called \emph{deconvolution problem}. (otherwise if the PSF is not determined we call the problem \emph{blind deconvolution problem}\cite{ayers1988}, which is not discussed in this article).

\subsection{Deconvolution methods and constraints}
Under suitable conditions the Fourier transform of a convolution is the point-wise product of Fourier transforms, stated by the convolution theorem. According to this direct restorations through Fourier transforms are widely used because the thoughts are straightforward and the restorations are non-iterative which usually guarantees the efficiency\cite{harris1966,ayers1988,zaroubi1995,loefdahl2007}. The presence of noise in the convolution equation \ref{eq-conv} gives the problem a statistics aspect. Therefore an estimate of the object intensity is also a solution of the deconvolution problem. If the object $O$ can be approximated by a simpler function with a few parameters, parametric fits are the best methods\cite{puetter2005}. For example the CLEAN algorithm, which is designed for radio imaging\cite{hoegbom1974}, approximates the object by a series of point sources, each with a pair of parameters i.e. the location and flux, and estimates the parameters through iterations. Otherwise the deconvolution problem must be modeled non-parametrically. In this case different statistics are used to approach an estimate of the object intensity, or a restoration. Landweber method minimizes the difference between the real observation $I$ and the convolution of the PSF $P$ and an estimate of the object $O$ uniformly\cite{landweber1951}. It implies a Gaussian model since this method gives a \emph{least-squares estimate}. With a Poisson model where the observation $I$ follows a Poisson distribution and the major contribution to the noise $N$ comes from photon noise, the Richardson-Lucy algorithm computes the \emph{maximum likelihood} (ML) estimates of the object intensity $\hat{O}$ iteratively\cite{richardson1972,lucy1974}. While in Bayesian statistics the \emph{maximum a posteriori} (MAP) algorithm maximizes the conditional probability density of the object intensity to achieve a MAP estimate\cite{starck2002}. However, methods introduced above do not produce uncertainty, but only a single restoration\cite{esch2004}. The expectation through Markov Chain Monte Carlo (EMC2) method as an extension of the Richardson-Lucy algorithm uses the same likelihood of the statistical model but sampling-based fitting method to produce uncertainty information\cite{esch2004}.

Deconvolution problem is known as an ill-posed problem\cite{byrne1998}. Constraints are vital for a desired solution of the problem -- an image represents the object with acceptable errors. Tighter constraints prevent underfitting the observation while looser constraints prevent overfitting it\cite{puetter2005}. The use of constraints is termed regularization\cite{lannes1987,hansen1994,starck2002}. Commonly used constraints include positivity, backgrounds, lower bounds (i.e. backgrounds) and upper bounds\cite{li1993}, band-limited constraints in Fourier domain (as various window functions or filters e.g. Wiener filter), imaging entropy, etc. Positivity constraint is implied in Richardson-Lucy algorithm as well as in the MAP algorithm. It helps Richardson-Lucy algorithm avoid amplifying noise, which is an important drawback of this algorithm\cite{dey2006}. In this article we will focus on background constraints in MAP deconvolution.

\subsection{About this article}
The concept of background constraint is straightforward -- the background of an image implies the lower bounds of intensities of pixels on the image. In general cases the background is estimated before it is used as a constraint.

On one hand the performance of deconvolution is substantially affected by the estimation of the background\cite{vankempen2000}. It is worth imposing prior constraints very cautiously because a regularization -- the use of constraints -- makes a compromise between ill-posedness and the accuracy of the deconvolution problem\cite{wing1991}. More restriction brings not only greater stability but also greater chance to eliminate correct solutions to the image reconstruction\cite{puetter2005}. For instance, enforcing a background constraint higher than the real one could cost a faint source significant loss or even lose the faint source, while providing a lower background constraint could make unpredictable artificial structures arise from noise, as demonstrated in the simulation section of this article.

On the other hand deconvolution methods introduced above are degraded because of the lack of uncertainty information in their results\cite{esch2004}. In other words, they produce point estimate of the unknown object such as a least-squares estimate, an ML estimate, or an MAP estimate instead of an interval estimate, which is however often interesting in astromony and astrophysics. The presence of an interval estimate, i.e., an image as the ``best estimate'' of the object with its uncertainty, will help discover faint sources, measure significance of restored structures, and retrieve rich statistics about interesting sources e.g. the fluxes and positions as well as the corresponding confidence intervals.

We suggest using probabilistic background constraints in deconvolution to take the uncertainty of the constraint into account rather than to indulge the uncertainty of it. Moreover, it is a crude yet refinable approach to the interval estimate of the deconvolution problem.

We elaborate the use of probabilistic background constraints in Sect. 2. Next a simulated example is presented in Sect. \ref{sect-simulation} and is discussed in Sect. 4, where we demonstrate how the approach we proposed help discover faint sources, measure significance of sources and fetch interesting statistics. Finally the conclusions are given in Sect. 5.


\section{Interval estimate with probabilistic background constraints}
\label{sect-theory}
\subsection{Uncertainties and interval estimate in deconvolution}
Many factors bring uncertainties to the deconvolution result in astronomical image restoration: uncertainty of the PSF, fluctuation of photon arrival i.e. the photon noise, the detector noise (e.g., as in the case of typical CCD detectors, the readout noise, the dark current, cosmic rays, uniformity of response, and bad pixels, etc.)\cite{mclean2008} and so on. So it makes sense to model the major noise and employ statistical methods in deconvolution. However the selections between different noise models and different methods add more uncertainties. Finally even with a proper noise model and a suitable deconvolution method there usually are still several adjustable parameters such as stopping conditions in iterative methods, or various prior constraints. Therefore uncertainties are inevitable in deconvolution. We target on an interval estimate of the unknown object in astronomical image deconvolution. Thus uncertainties are limited in confidence intervals.

\subsection{Approach to an interval estimate: monte-carlo experiment and marginal distributions}
\label{sect-approach}
In general cases monte-carlo experiments are feasible to calculate an approximate distribution of a given random variable. What we want to calculate is the distribution of possible images given the observed data and uncertain background constraints. And from such distribution we can find some ``best estimate'' of the image as well as the corresponding confidence interval of it. When taking an image as a multivariate random variable, which groups image value on each pixel together as individual variables, the image space is so huge, hence most of the draws sampled from this space are not likely to permit the given observed data, thus contributions to the distribution from them are quite so small, so monte-carlo experiment sampling directly from the space is largely impractical.

Instead of directly from the image space, we generate draws from solutions of statistical methods, e.g., MAP estimates. For each draw we calculate its probability and use this probability to weigh its occurrence. On one hand, in view of the reduction of the image space, the extra calculations of probabilities have the merits. On the other hand, given the observed data, the solution converges towards the MAP estimate or ``best estimate'' of other statistics among all possible estimates permitted by the enforced constraints. More specifically, the solution usually converges towards the \emph{mode}, given a specific probability density function of each estimate satisfying the constraints, i.e., estimates with similar probability densities are close to each other. However estimates with similar but different background constraints are not that close to each other. Thus we believe the distribution of these solutions summarizes the distribution of the population (i.e., the set of all possible estimates permitted by given constraints) well enough.

The ``goodness'' of an estimate calculated with statistical methods such as the likelihood function or the posterior probability is related to joint probability density function of the estimate, i.e., the image, but not directly related to image values on individual pixels. However we usually pay attention to distributions of image values of interesting structures or even specific pixels rather than the image as a whole. So the interval estimate we need is not for the whole image but for image values of individual pixels. To achieve that we calculate marginal distributions of image values on the pixels.

\subsection{Probabilistic background constraints in deconvolution}
\label{sect-theory-background-estimate}
In astronomical image deconvolution the background arises from both the remote astrophysical components e.g. far away galaxies, diffuse sources, etc., and the local instrumental factors. Compared with PSF, detector noise, and other uncertainties sources the background is more difficult to be determined precisely through prior calibrations in a single observation. This article focuses on the background constraints only. Here we will not discuss uncertainties from PSF, noise, or other aspects.

While we estimate the background from the observed data we can also estimate the confidence interval of the estimated background. The estimation can be achieved by following steps. First, extract the observed data dominated by background luminosities. We call it background data. Second, according to the background data itself or prior knowledge select a proper background model e.g. a spatial function with several adjustable parameters. Finally, calculate the ``best-fit'' parameters as well as their confidence intervals with the background data. 

For example, If the object consists of sparse sources only and the background has only large-scale spatial structures, the background can be approximated with a simple function or even a constant soon after the sources have been removed from the observed data. Clipping methods\cite{vankempen2000,starck2006} and CLEAN algorithm\cite{li1993} are both workable to extract the background data. We pick CLEAN algorithm and assume the image values on the residual map $I_R(x)$, which serves as the extracted background data, follows a normal distribution. As a result we can use least-squares method to estimate the background parameters as well as their confidence intervals. To make the example as simple as possible we assume further that the background does not have spatial structures at all so we can use a constant $B$ to model the background. Thus the average of the background data $\overline{I_R(x)}$ is exactly the ``best-fit'' background, i.e. $\hat{B}=\overline{I_R(x)}$ (since the convolution of a constant and a normalized function is the constant itself, i.e., $(B \ast P)(x) = B$) and the variance of the background data $Var[I_R(x)]=\overline{\left(I_R(x)-\overline{I_R(x)}\right)^2}$ reflects its uncertainty. Given the \emph{central limit theorem} the estimated background follows a normal distribution approximately, and the mean of the estimated background is also $\hat{B}$ while its standard deviation is related to the variance of the background data as $\sigma_{\hat{B}}=(Var[I_R(x)]/N)^{1/2}$ given the background data consists of $N$ observed count rates.

In iterative methods such as Richardson-Lucy algorithm the background constraint is enforced as:
\begin{equation}
\hat{O}'(x) =
  \begin{cases}
  \hat{O}(x) & \hat{O}(x) \geq B \\
  B          & \text{otherwise}
  \end{cases}\text{,}
\label{eq-background-constraint}
\end{equation}
where $\hat{O}(x)$ is an estimate of the unknown object thus an image at a iteration step, $\hat{O}'(x)$ is the image corrected by the background constraint\cite{vankempen2000,starck2002}. $B$ usually is the estimated background. We suggest using a probabilistic background constraint $B_k$ sampled from the distribution of the background estimate through a series of iterations, namely, performing the $k$-th deconvolution with the background constraint $B_k$. This is the monte-carlo experiment we suggest in Sect. \ref{sect-approach}.

\section{Simulation and results}
\label{sect-simulation}
\subsection{Configuration of simulation}
\subsubsection{Object, PSF and observed data}
\label{sect-config}
\begin{figure}[t]
\centerline{\includegraphics[width=\linewidth]{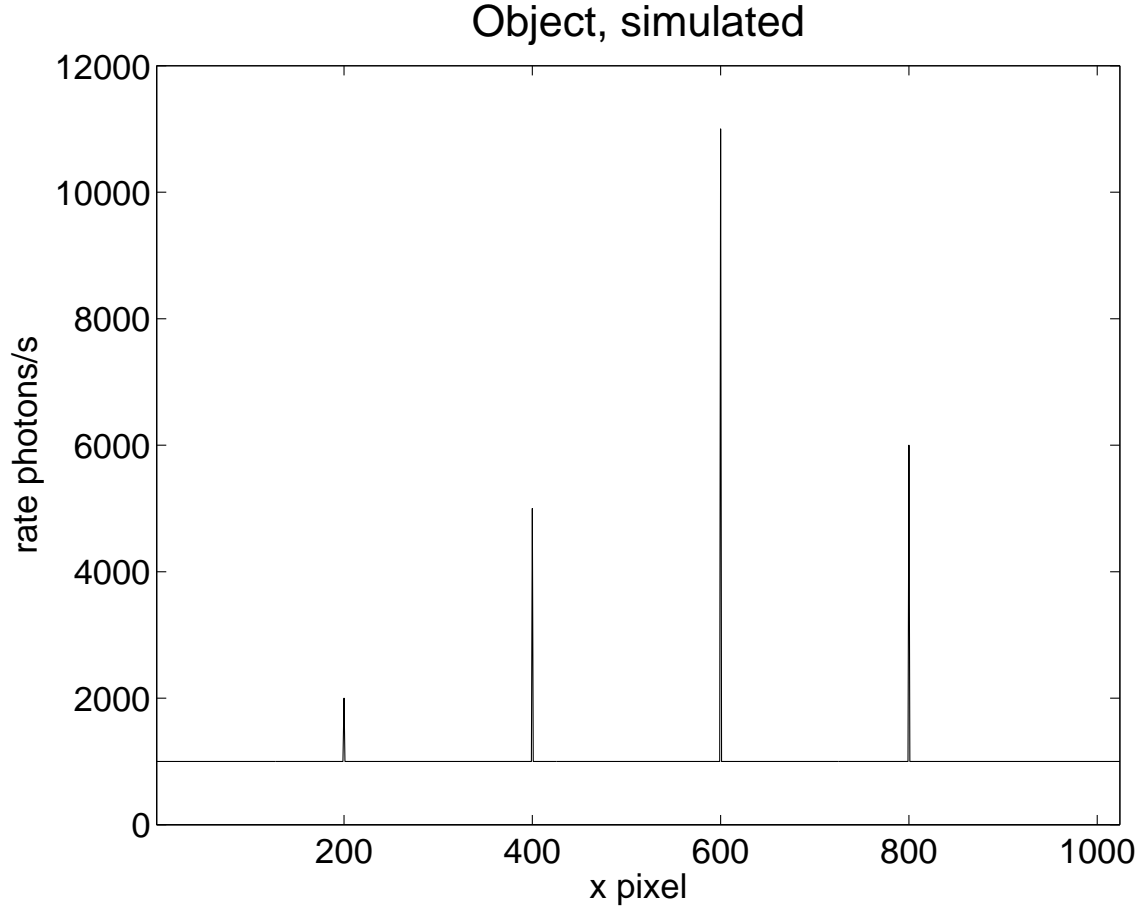}}
\caption{Simulated object $O(x)$. Sources: $1\,000\;photons/s$ at $200\;pixel$, $4\,000\;photons/s$ at $400\;pixel$, $10\,000\;photons/s$ at $600\;pixel$, and $5\,000\;photons/s$ at $800\;pixel$. The intensity of the background is $1\,000\;photons/s$.}
\label{fg-obj}
\centering
\end{figure}

The simulated object consists of several point sources and the background are shown in Fig. \ref{fg-obj}.

\begin{figure}[t]
\centerline{\includegraphics[width=\linewidth]{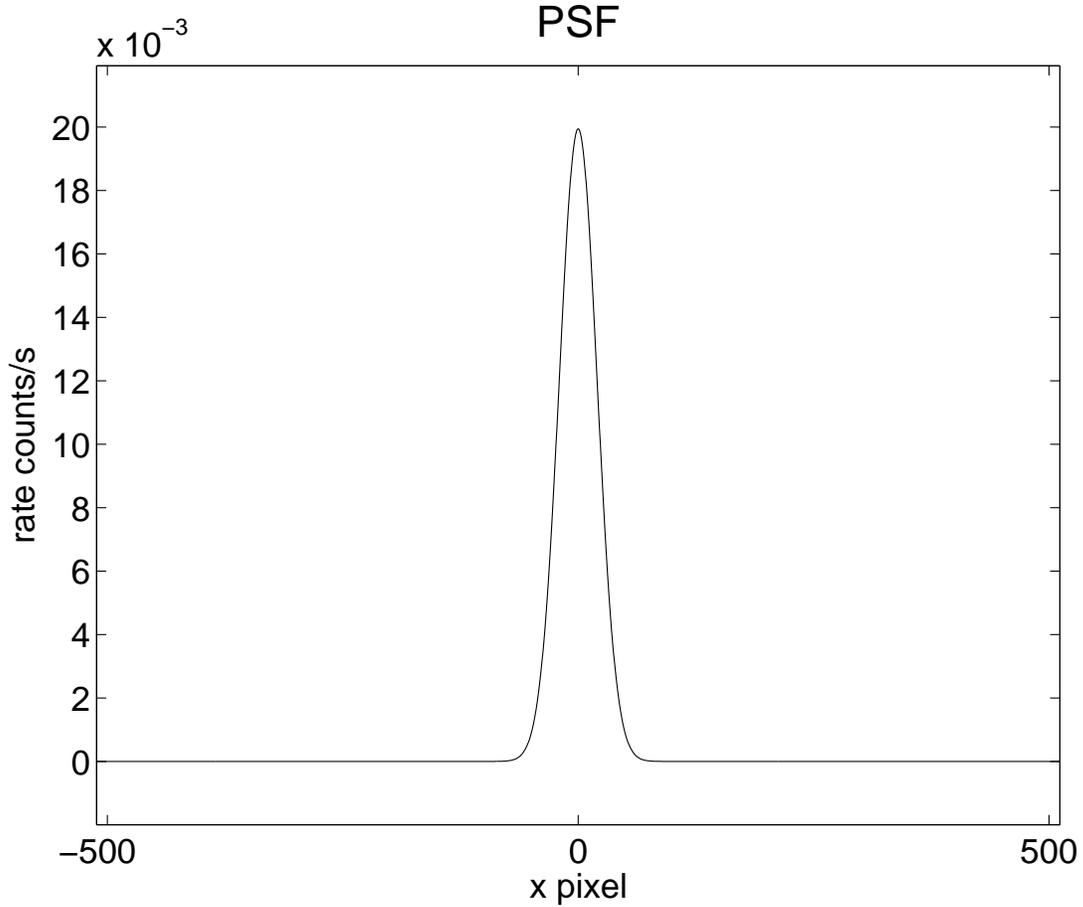}}
\caption{Simulated PSF $P(x)$.}
\label{fg-psf}
\end{figure}

The PSF is simulated with a normalized Gaussian bell-curve with FWHM (full width at half maximum) of $46\;pixels$ (as Fig. \ref{fg-psf}).

\begin{figure}[t]
\centerline{\includegraphics[width=\linewidth]{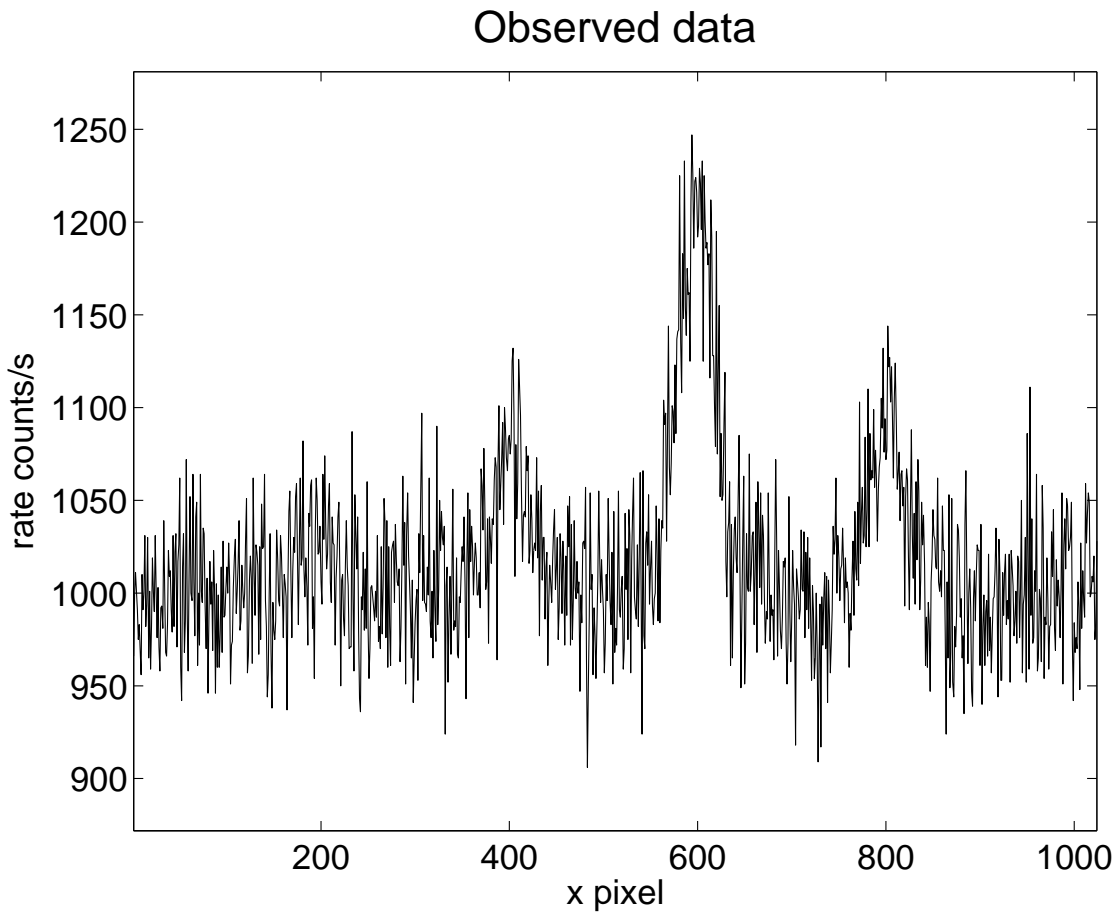}}
\caption{Simulated observed data.}
\label{fg-obs}
\end{figure}

The observed data (e.g. Fig. \ref{fg-obs}) is simulated by:
\begin{enumerate}
\item convolving $O(x)$ with $P(x)$ (let $I(x) = (O \ast P)(x)$),
\item adding photon noise to $I(x)$ (we use normal fluctuation $N(0,I(x))$ to approximate the photon noise with Poisson distributions, $Pois(\sqrt{I(x)})$.),
\item adding additive normal noise $N_{obs}$ ($N_{obs} \sim N(0,25)$) to $I(x)$ to simulate other noise sources, and
\item rounding each observed data value to its nearest integer to simulate the round-off error.
\end{enumerate}
The simulated object as well as the PSF and the observed data are discretized into $1\,024 \times 1\;pixels$ vectors. Thus the coordinate variable $x$ of either $O(x)$, $P(x)$ or $I(x)$ is discretized into a dimensionless integer variable in the sense that it denotes an arbitrary pixel.

\subsubsection{Background estimation}
\label{sect-sim-bgest}
\begin{figure}[t]
\centerline{\includegraphics[width=\linewidth]{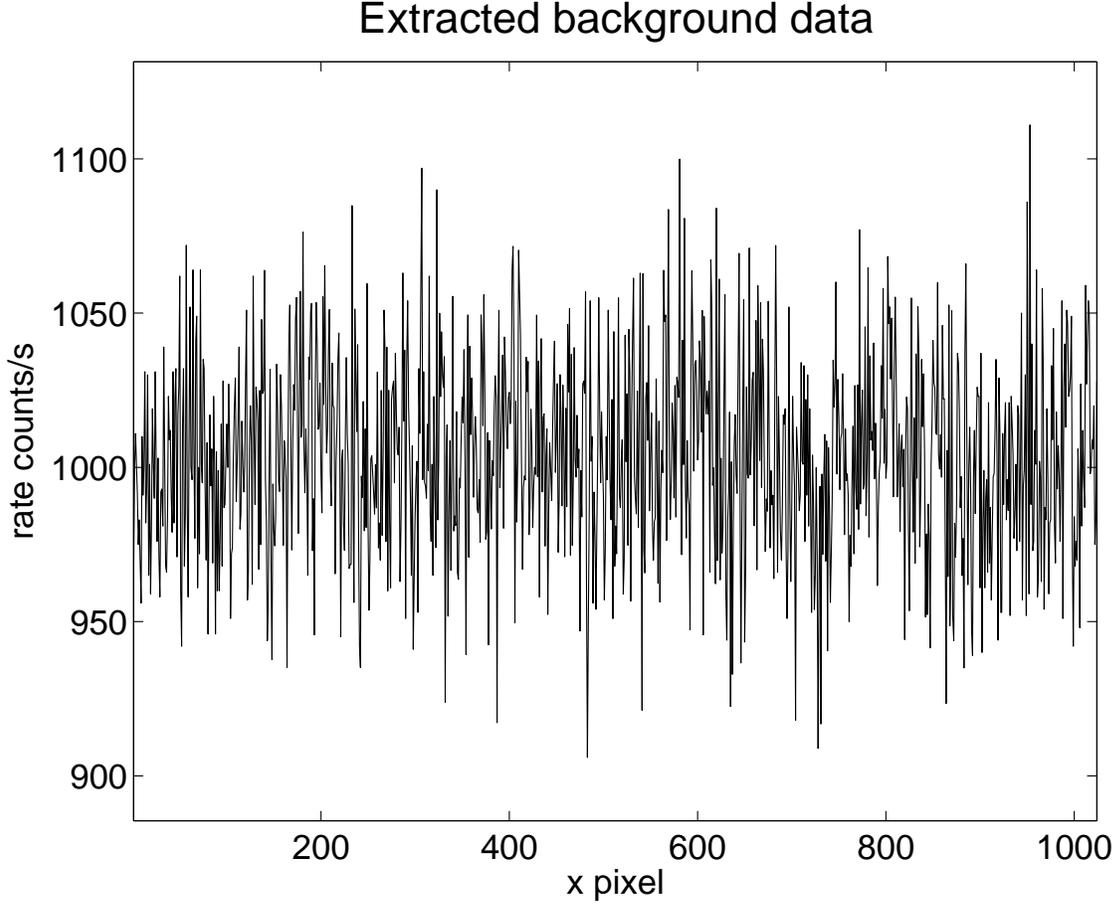}}
\caption{Background data $I_R(x)$.}
\label{fg-bg}
\end{figure}

CLEAN algorithm is employed as introduced in Sect. \ref{sect-theory-background-estimate}. The residual map, i.e. the background data is extracted as shown in Fig. \ref{fg-bg}.
The arithmetic mean of the data $\overline{I_R(x)} = 1\,002.1\;counts/s$ while its standard deviation $\sigma_{I_R(x)} \approx \Bigg[\overline{(I_R(x) - \overline{I_R(x)})^2}\Bigg]^{1/2} = 31.5\;counts/s$.

From the background data $I_R(x)$ (Fig. \ref{fg-bg}) we obtain:
\begin{itemize}
\item the ``best estimate'' of the background intensity $\hat{B} = 1\,002.1\;counts/s$, and
\item the estimate of the background is approximately distributed as $N(1\,002.1, 1.2)$.
\end{itemize}

\subsubsection{MAP deconvolution}
\begin{figure}[t]
\centerline{\includegraphics[width=\linewidth]{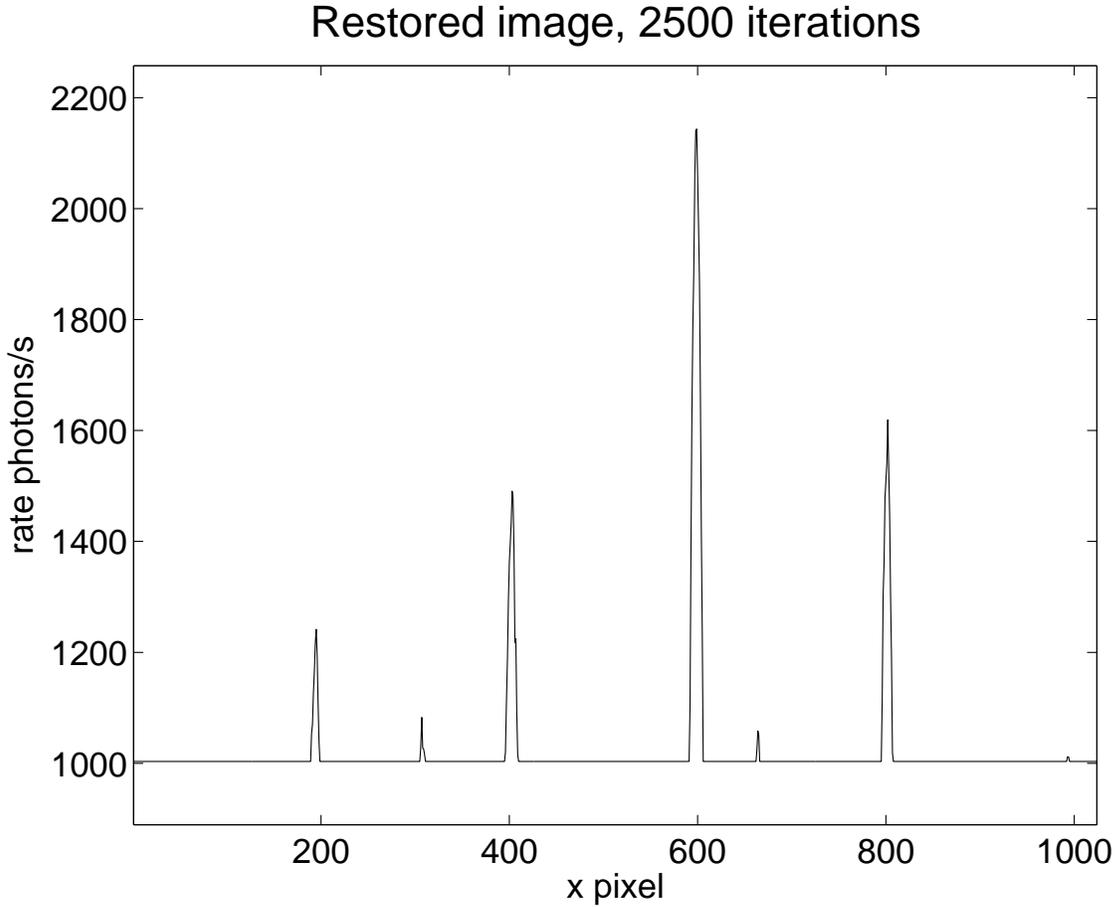}}
\caption{Object estimate $\hat{O}_{MAP}$. Deconvolving with $2\,500$ MAP iterations, using $\hat{B} = \overline{I_R(x)}$ as the background constraint.}
\label{fg-map}
\end{figure}

We employ MAP method to solve the deconvolution problem. In this configuration we find within $2\,500$ iterations the maximization process converges well enough to provide a stable estimate, i.e. $\hat{O}_{MAP}$, as shown in Fig. \ref{fg-map}.

\begin{figure}[t]
\centerline{\includegraphics[width=\linewidth]{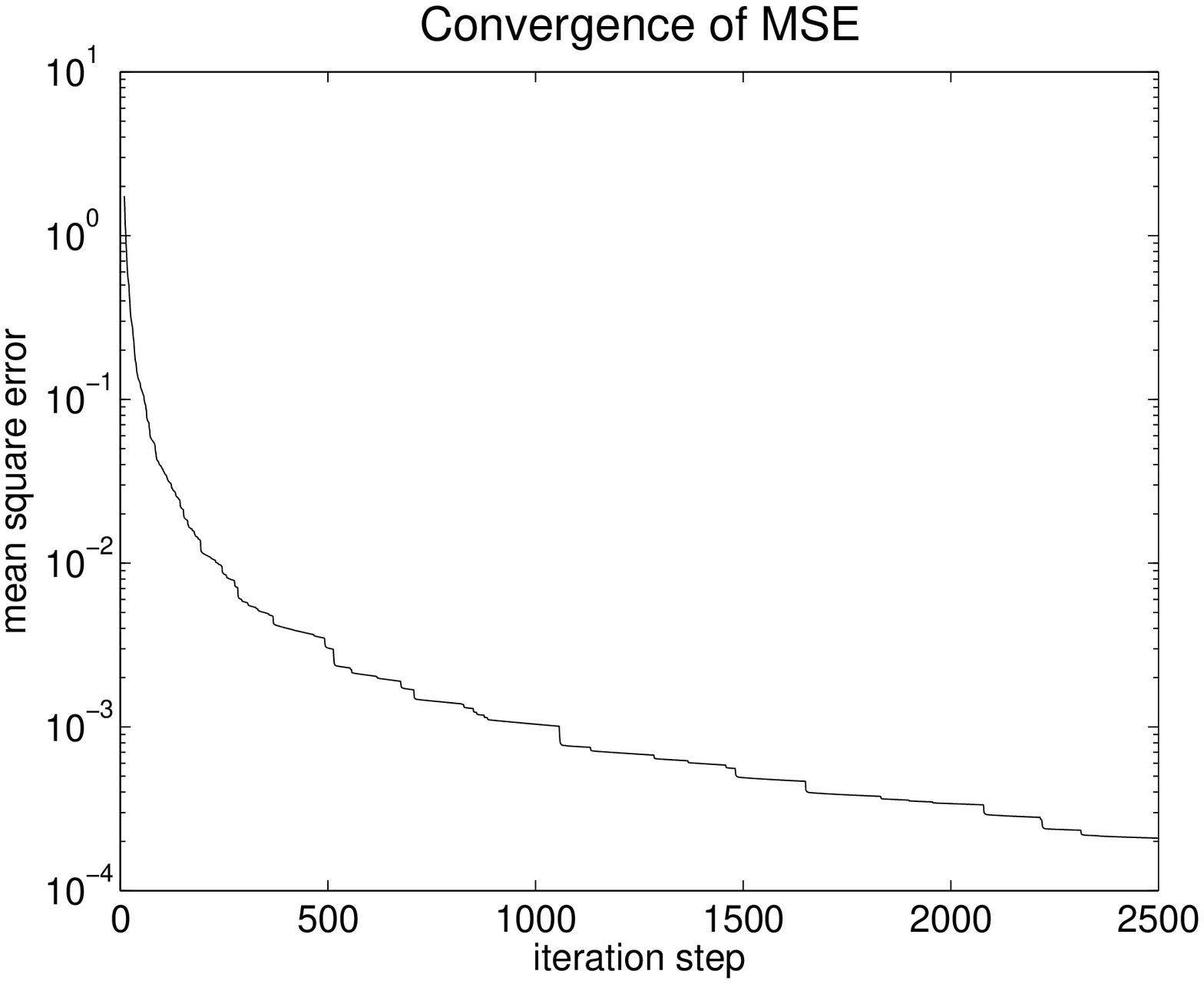}}
\caption{Convergence of MSE (mean square error). It shows the MSE between two adjacent iteration steps.}
\label{fg-converge-mse}
\end{figure}

\begin{figure}[t]
\centerline{\includegraphics[width=\linewidth]{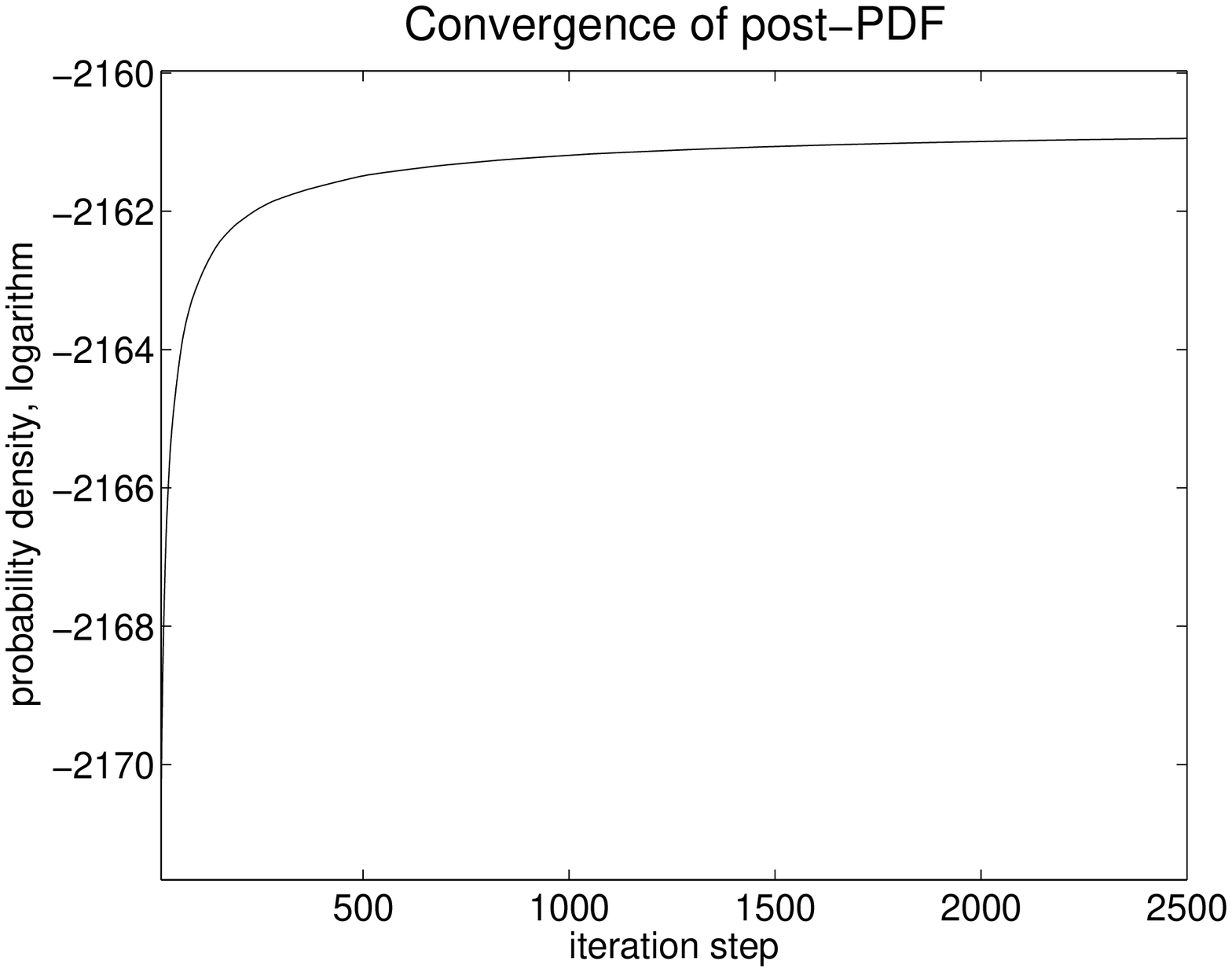}}
\caption{Convergence of posterior PDF (probability density function). It shows the logarithm to base $10$ of the posterior probability density function of the estimate in each iteration step.}
\label{fg-converge-prob}
\end{figure}

We also investigate the convergence of this method by observe the mean square error as well as the posterior probability density going with the iterations. The mean square error in the $k$-th step is defined as
\begin{equation*}
\epsilon^{(k)}=\overline{(O_{MAP}^{(k)}(x)-O_{MAP}^{(k-1)}(x))^2}
\end{equation*}
and it measures the difference between two adjacent estimates, say, the $k$-th estimate and its previous one. The posterior probability density is defined as\cite{starck2002}
\begin{equation}
p(O_{MAP}^{(k)}|I)=\Pi_x\frac{(O_{MAP}^{(k)} \ast P)^{I(x)}(x) e^{-(O_{MAP}^{(k)} \ast P)(x)}}{I(x)!}
\label{eq-postprob}
\end{equation}
if assume a uniform prior probability of the unknown object. It is convenient calculating the logarithm to base $10$ of the probability density in numerically since even that of the most probable estimate is quite small. See Fig. \ref{fg-converge-mse} and Fig. \ref{fg-converge-prob} for the mean square error and posterior probability density respectively.

The posterior probability density shown in Equation \ref{eq-postprob} is evaluated approximately.

\subsection{Monte-carlo experiment and results}
\subsubsection{Monte-carlo experiment}
As suggest in Sect. \ref{sect-approach} we setup a monte-carlo experiment to calculate marginal distributions of image values on individual pixels. The experiment consists of following steps:
\begin{enumerate}
\item Generate a random draw $B_k$ from $N(1\,002.1, 1.2)$, which is the distribution of the background estimates.
\item Use $B_k$ as the background constraint in deconvolving the observed data $I(x)$.
\item As a result we get a MAP solution $\hat{O}_k(x)$ and its probability $p(\hat{O}_k|I)$.
\item Repeat from 1 to 3 for $K$ times, store all $K$ $\hat{O}_k(x)$ and $p(\hat{O}_k|I)$.
\end{enumerate}
See the diagram of monte-carlo experiment below for the setup.
\begin{displaymath}
\begin{diagram}
  \node[2]{\text{Distribution of background estimates}} \arrow{s,r}{\text{sample}}\\
  \node[2]{\hat{B}_k} \arrow{s,r}{\text{regularize}}\\
  \node{I} \arrow{nne,t}{\text{extract and estimate}} \arrow{e,e} \node{\text{MAP method}} \arrow{e,e} \node{\hat{O}_{k},\;p(\hat{O}_k|I)}
\end{diagram}
\end{displaymath}
In the diagram $I$ is the observed data, $\hat{O}_{k}$ and $p(\hat{O}_k|I)$ are the image and its probability given the observed data in 
the $k$-th deconvolution with MAP method.

\subsubsection{Marginal distribution and interval estimate}
\begin{figure}[t]
\centerline{\includegraphics[width=0.6\linewidth]{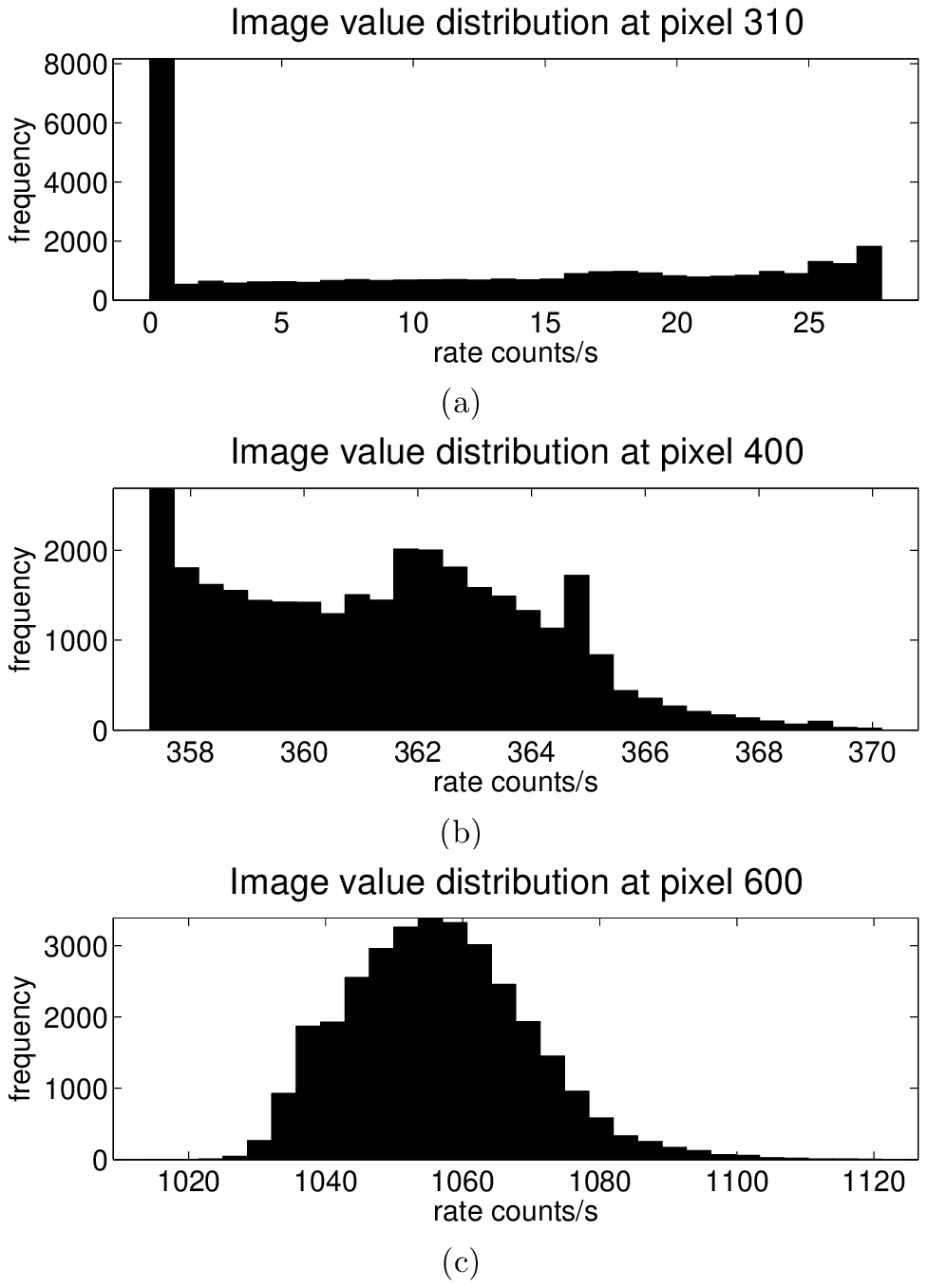}}
\caption{Normalized occurrence frequency of image values (backgrounds are truncated) on $x=310\;pixel$ (top), $x=400\;pixel$ (middle), and $x=600\;pixel$ (bottom).}
\label{fg-ratedist}
\end{figure}

First we calculate the histogram for image value on each pixel. Then we calculate the total posterior probability density for each interval of each histogram. Finally we scale the bar on each interval by the total posterior probability density to update the histograms. See Fig. \ref{fg-ratedist} for examples. Such histograms portray the marginal distributions of image values individually.

Because the background is not restored by deconvolution process but enforced by the constraint, it is truncated from each image value in Fig. \ref{fg-ratedist}.

\begin{figure}[t]
\centerline{\includegraphics[width=\linewidth]{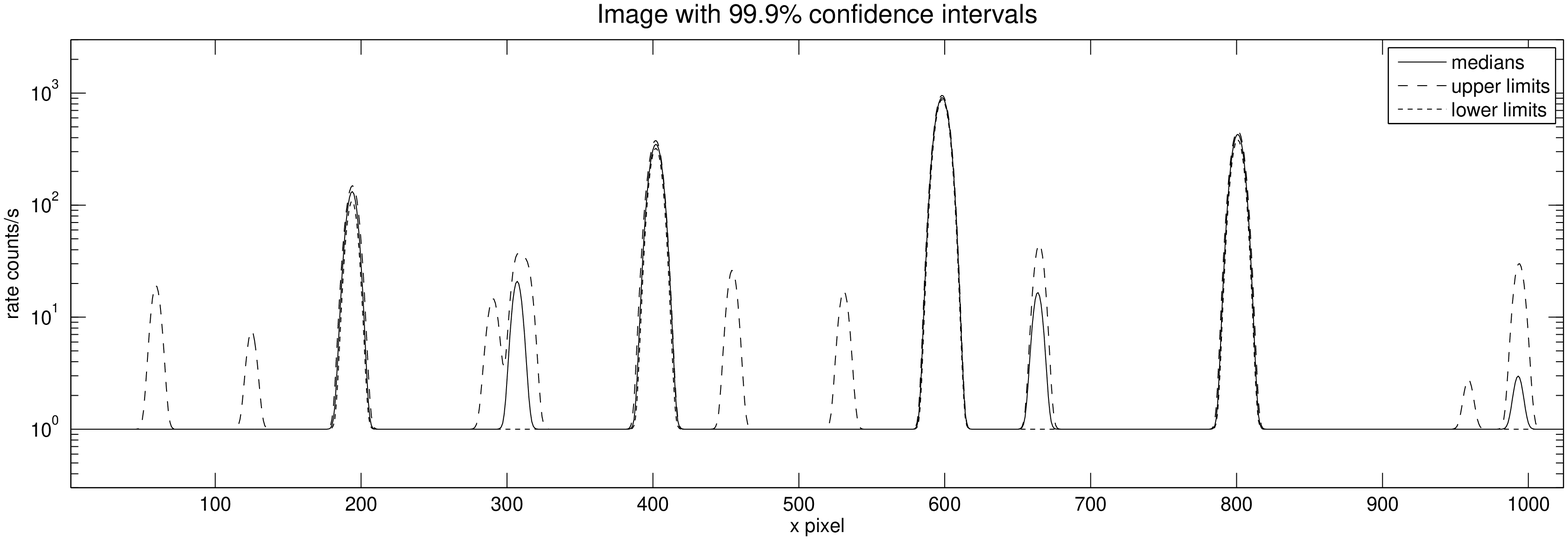}}
\caption{Medians of marginal distributions of image values for all the pixels as well as their $99.9\%$ (about $3.3\sigma$) confidence intervals in images. Medians, upper limits and lower limits are indicated by solid line, dashes, and dots respectively. Data series are shifted, smoothed and plotted with logarithmic scaled $y$-axis to highlight differences between different series especially on faint structures.}
\label{fg-median}
\end{figure}

We find the median as well as the central confidence interval (CI) of the marginal distribution of the image value for each pixel from its histogram. For example there are $0.05\%$ image values are greater than the upper limit of the $99.9\%$ confidence interval and $0.05\%$ image values are less than the lower limit of it. See the medians as well as the confidence intervals in Fig. \ref{fg-median}.

\subsubsection{Source retrieval and significance level}
\label{sect-flux}
\begin{figure}[t]
\centerline{\includegraphics[width=\linewidth]{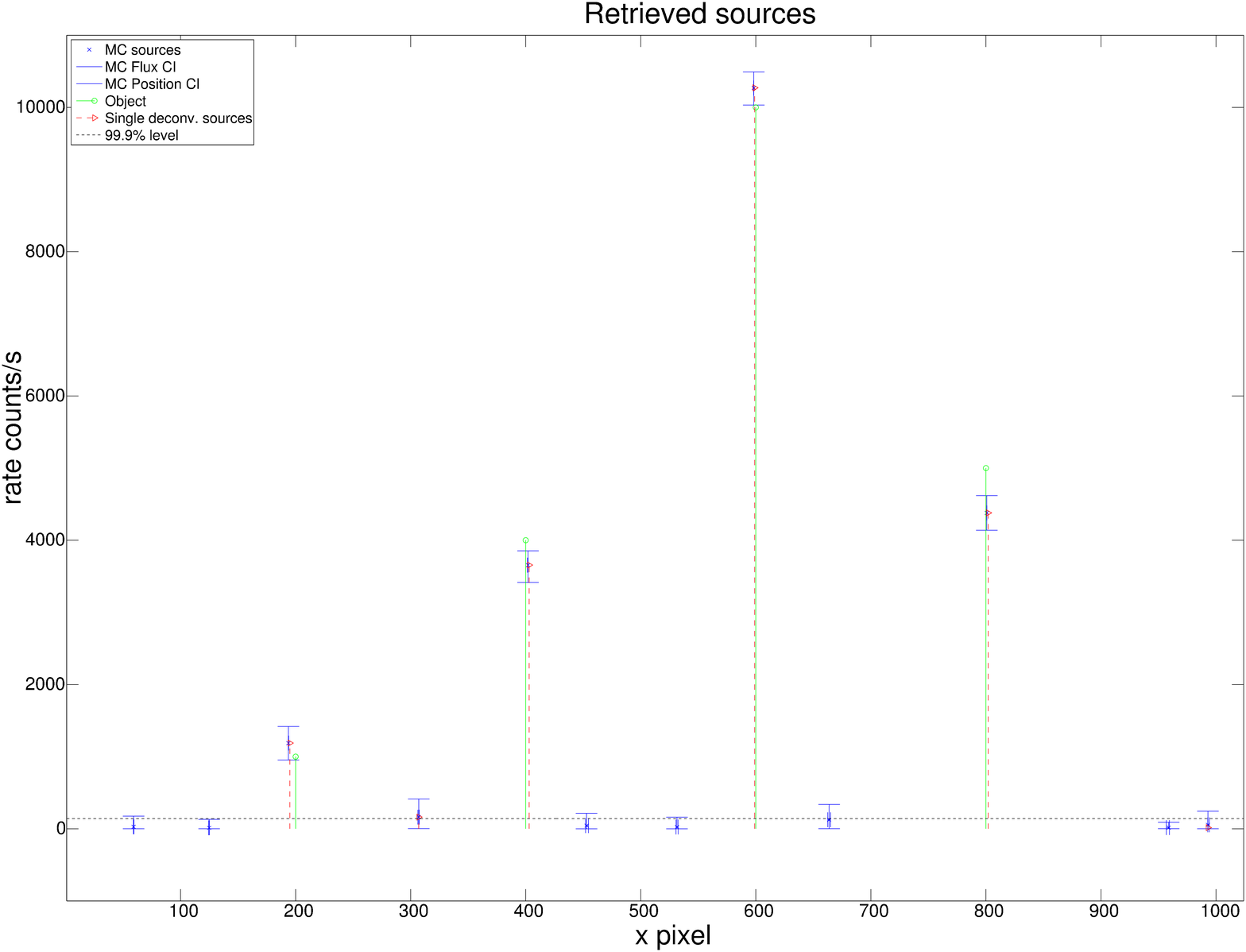}}
\caption{(Color online) Fluxes ($y$-coordinates) and positions ($x$-coordinates) of retrieved sources, $99.9\%$ confidence intervals of both fluxes and positions, as well as the $0.01\%$ significance level for fluxes of retrieved sources. Blue crosses indicate the fluxes and of sources retrieved from monte-carlo experiment using probabilistic background constraints, while error bars along $x$-axis indicate central $99.9\%$ confidence intervals of positions and error bars along $y$-axis indicate central $99.9\%$ confidence intervals of fluxes. Green solid lines with circles indicate sources in the simulated object. Red dashed lines with triangles indicate sources retrieved from a single run of MAP deconvolution. Black dotted line indicates the $0.01\%$ significance level. Here the significance level is $142\;counts/s$.}
\label{fg-flux-pos}
\end{figure}

We use a program to search for continuous structures in each image restored from the monte-carlo experiment while marginal distributions of image values are being calculated. Each structure is considered as a source. We take the sum of all the image values of pixels within the range of a source as its flux, while we take the \emph{barycentric position}, weighted by image values, as its position. We account sources overlap with one another between different images the same source. So we obtain a distribution for the flux of each source and that for its position. From the distribution, we can not only calculate the ``best estimate'' of each flux or position but also the corresponding confidence interval (as shown in Fig. \ref{fg-flux-pos}).

\begin{figure}[t]
\centerline{\includegraphics[width=0.6\linewidth]{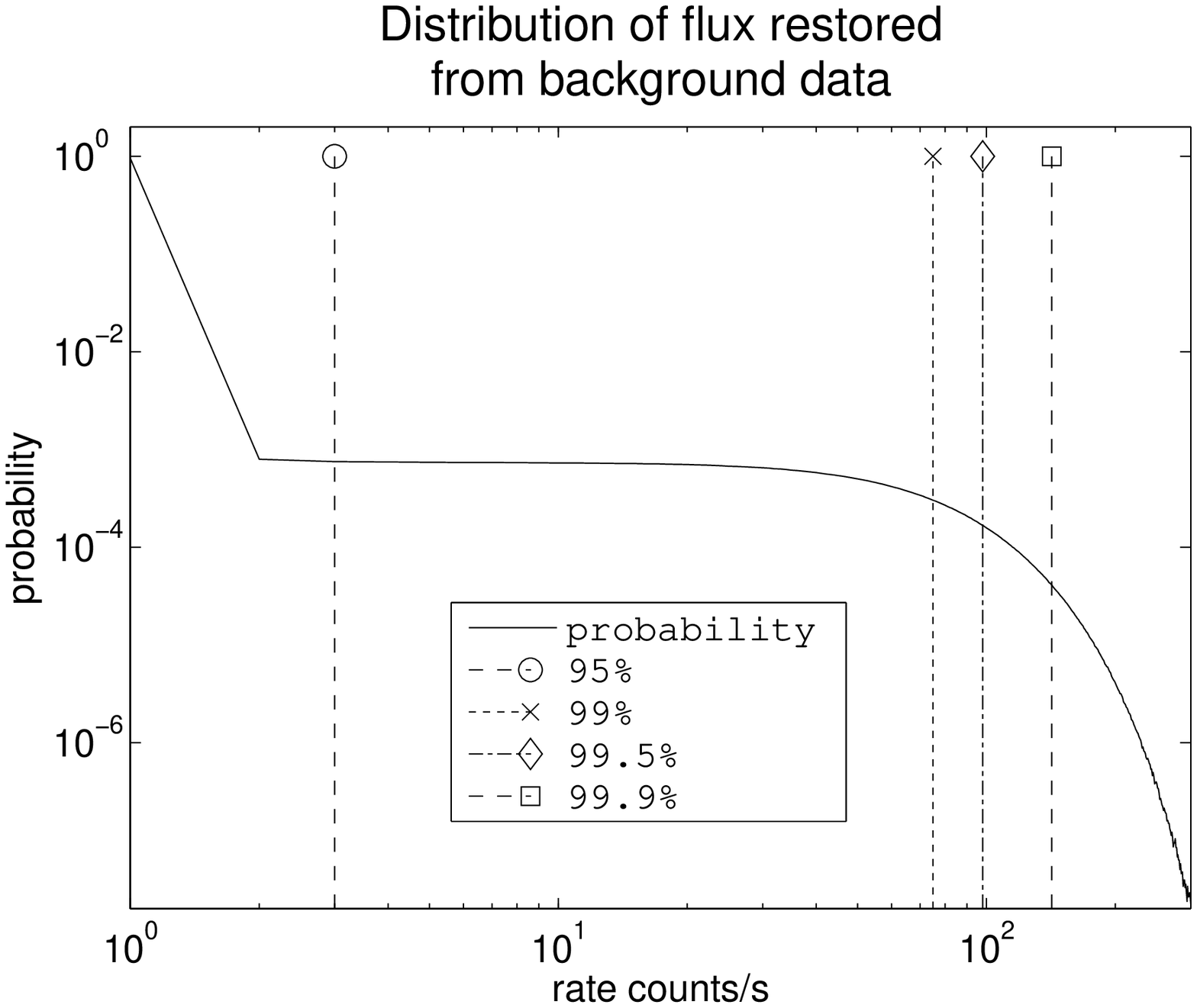}}
\caption{Null distribution of flux restored from random background data sets with the same means and variances as the background data extracted from the observed data. The dashed line with a circle, the dotted line with a cross, dash-and-dotted line with a diamond, and the dashed line with a square denote the $95\%$, $99\%$, $99.5\%$, and $99.9\%$ cumulative probabilities respectively.}
\label{fg-bgmc}
\end{figure}

We perform a monte-carlo experiment of deconvolutions on random background data sets to calculate the null distribution of flux in restored images. The random background data sets have the same means and variances as the background data extracted from the observed data in Sect. \ref{sect-sim-bgest}. We retrieve fluxes of ``sources'' arising from ``images'' restored from the background data. Of course these restored ``sources'' arise from magnified noise since there is no source in the background data at all. Finally we calculate the null distribution of fluxes of these artificial ``sources'' we restored from random background data sets. See the distribution in Fig. \ref{fg-bgmc}.

Vertical lines in Fig. \ref{fg-bgmc} indicate significance levels of sources restored from observed data. For example there are only $5\%$ artificial ``sources'' arising from the noise have fluxes greater than the flux where the cumulative probability is $95\%$. So the dashed line denoting this probability also indicates the $5\%$ (about $2\sigma$) significance level, i.e., $3\;counts/s$. The $1\%$ (about $2.6\sigma$), $0.5\%$ (about $2.8\sigma$), and $0.1\%$ (about $3.3\sigma$) significance levels are $80\;counts/s$, $98\;counts/s$ and $142\;counts/s$ respectively. 

\section{Discussion}
\subsection{About results of numerical simulation}
In Sect. \ref{sect-simulation} we calculated the ``best estimate'' of the unknown object as well as its confidence intervals as shown in Fig. \ref{fg-median}. The performance of deconvolution is affected by the background constraint we enforced. Compared with the object (as in Fig. \ref{fg-obj}), there are artificial structures in the restored images, and the intensities of these structures vary in images restored with different background constraints. In a real case we certainly have no idea whether such a structure in an image restored from observed data is artificial and due to the underestimate of the background, or a real yet faint source. Given the confidence interval, however, we know the lower bounds. So if the lower bound of a structure is still positive we know this structure is not due to the underestimate of the background at least. For example, as shown in Fig. \ref{fg-map}, there are artifacts in the image restored by a single run of deconvolution. They are mainly due to the underestimate of the background since their presences are sensitive to the background constraints obviously. For bright sources, their confidence intervals reveal uncertainties due to the estimate of the background.

As shown in Fig. \ref{fg-ratedist} the marginal distributions are highly skewed and some of them are bimodal. Compared with other measures of central tendency such as mean and mode, we use the median as the best estimate. The mode of such distribution converges only with plenty of scores been accumulated. The mean tends to be affected by extreme values. Both the mode and the mean can be out of the central confidence interval. We believe the median is good enough to measure the central tendency and it is insensitive to the sample size.

As shown in Fig. \ref{fg-flux-pos} the fluxes of sources are more dispersed than the positions. This is due to the limitation as well as the aim of the approach we take to calculate the confidence intervals. Our approach is aimed at and limited to the background constraints. So the confidence intervals are all about the background constraints. Therefore it is clear that the background constraint mainly affects the intensities of restored image values in MAP deconvolution. Also we see in Fig. \ref{fg-flux-pos} even $99.9\%$ confidence intervals do not cover all correct fluxes and positions. It is still because the intervals only measure the uncertainties brought by the background constraint. The extra errors are mainly caused by simulated photon noise and detector noise. Since we perform deconvolutions only on a single set of observed data, errors caused by these noises can not be eliminated.

\subsection{Feasibility of our approach in general cases}
In Sect. \ref{sect-theory} and and Sect. \ref{sect-simulation} we focus our attention on background constraint in deconvolution. In Sect. \ref{sect-simulation} we perform numerical simulations employing MAP estimate. However the approach suggest in this article is not confined in the case we elaborated or simulated.

With necessary prior analysis on the distribution of such as the error of PSF, the detector noise, etc., we can always perform monte-carlo experiment of deconvolutions with random draw of corresponding input. Then with the results we can calculate marginal distributions of image values on all pixels immediately. Although these already reflect the uncertainty brought by the corresponding input, since in our approach the results of monte-carlo experiment are stored individually instead of merged prematurely, interesting information can still be retrieved. For example we retrieve the fluxes and positions from the restored images in Sect. \ref{sect-flux}. It is remarkable that the retrieval procedure works as a blind parametric estimate -- it is not necessary to know where a source is or how many sources there are since it is easier to locate sources in restored images.

We also suggest running deconvolutions on background data to provide for significance levels of sources in images restored in observed data, as shown in Fig. \ref{fg-bgmc}. This is about another source of artifacts in image restoration -- magnified noise in observed data. We take only the noise on background in to account, however, artifacts could also arise nearby a bright source due to the noise on its flux spreaded by PSF. To analyze significance levels of source restored nearby a bright source, the bright source should also be put into the data.

\section{Conclusions}
We elaborated and simulated using probabilistic background constraints to build monte-carlo experiment of deconvolutions. From the results we calculated interval estimates of restored images that reveal the uncertainties brought by the background constraints to the deconvolution. The background constraints affect mainly the intensities of restored image values. This approach helps discriminate structures arising from restored images caused by the underestimate of the background from others and discover those disappear due to the overestimate of the background. We can also use this approach to analyze uncertainties from other aspects thus eliminate errors from other aspects.



\section*{Acknowledgements}
This work was supported by the National Natural Science Foundation of China (NSFC) under Grant No. 11173038, and also by Tsinghua University Initiative Scientific Research Program under Grant No. 20111081102.

\end{document}